\documentclass[11pt,english,a4paper,floatfix]{article}
\usepackage{jcappub}

\pdfoutput=1
\usepackage{bm}
\usepackage{amsfonts}
\usepackage{subfigure}

\newcommand{\be}{\begin{equation}}
\newcommand{\ee}{\end{equation}}
\newcommand{\bea}{\begin{eqnarray}}
\newcommand{\eea}{\end{eqnarray}}

\renewcommand{\vec}[1]{\boldsymbol{#1}}

\usepackage{lmodern}
\usepackage{slantsc}
\newcommand{\CONCEPT}{\textsc{co\textsl{n}cept}}
\newcommand{\CLASS}{\textsc{class}}

\newcommand{\appropto}{\mathrel{\vcenter{
  \offinterlineskip\halign{\hfil$##$\cr
    \propto\cr\noalign{\kern2pt}\sim\cr\noalign{\kern-2pt}}}}}

\usepackage[scr=boondox]{mathalfa}

\usepackage{mathtools}

\usepackage{siunitx}

\begin{document}

%%%%%%%%%%%%%%%%%%%%%%%%%%%%%%%%%%%%%%%%%%%%%%%%%%%%%%%%%%%%%%%%%%%%%%
% Frontpage %%%%%%%%%%%%%%%%%%%%%%%%%%%%%%%%%%%%%%%%%%%%%%%%%%%%%%%%%%
%%%%%%%%%%%%%%%%%%%%%%%%%%%%%%%%%%%%%%%%%%%%%%%%%%%%%%%%%%%%%%%%%%%%%%

\title{Fully relativistic treatment of decaying cold dark matter in $N$-body simulations}

\author[a]{Jeppe Dakin,}
\author[a]{Steen Hannestad,}
\author[b]{Thomas Tram}

\affiliation[a]{Department of Physics and Astronomy, Aarhus University,
 DK-8000 Aarhus C, Denmark}
\affiliation[b]{Aarhus Institute of Advanced Studies (AIAS), Aarhus University, DK--8000 Aarhus C, Denmark}

\emailAdd{dakin@phys.au.dk}
\emailAdd{sth@phys.au.dk}
\emailAdd{thomas.tram@aias.au.dk}

\abstract{
We present $N$-body simulations in which either all, or a fraction of, the cold dark matter decays non-relativistically to a relativistic, non-interacting dark radiation component. All effects from radiation and general relativity are self-consistently included at the level of linear perturbation theory, and our simulation results therefore match those from linear Einstein-Boltzmann codes such as \CLASS{} in the appropriate large-scale limit. We also find that standard, Newtonian $N$-body simulations adequately describe the non-linear evolution at smaller scales ($k \gtrsim 0.1 \, h/{\rm Mpc}$) in this type of model, provided that the mass of the decaying component is modified correctly, and that the background evolution is correctly treated. That is, for studies of small scales, effects from general relativity and radiation can be safely neglected.
}

\maketitle

%%%%%%%%%%%%%%%%%%%%%%%%%%%%%%%%%%%%%%%%%%%%%%%%%%%%%%%%%%%%%%%%%%%%%%%%%%%%%%%%%%%%%%%%%%%%%%%%%%

\section{Introduction}

The dark matter component of the Universe is usually assumed to be stable and interact only gravitationally. Such a component generally provides a very good fit to observations.
However, given that we do not know the true nature of dark matter, it is worthwhile to investigate its possible non-gravitational interactions.

One possibility which has been extensively studied is that dark matter possesses self-interactions, typically in the form of exchange of some new vector boson (\cite{Carlson:1992fn,Spergel:1999mh}, see e.g.\ \cite{Tulin:2017ara} for a recent review).
This leads to the possibility of rapid dark matter scatterings in regions of high density, which in turn can affect the formation of bound structures such as galaxies.

Another possibility is that dark matter is unstable. If all dark matter can decay, there are already quite stringent bounds on its lifetime. This possibility was (to our knowledge) first investigated in the pioneering paper \cite{Flores:1986jn}.
In dark matter decay it is important to distinguish between models in which the dark matter decays into standard model particles and models where dark matter decays to other particles in the dark sector. In the former case there are extremely stringent bounds on the lifetime because dark matter decay inevitably lead to energy injection into the electromagnetically interacting standard model plasma, typically in the form of either photons and/or electron-positron pairs (see e.g.\ \cite{Ibarra:2013zia,Dugger:2010ys,Cirelli:2009dv,Bertone:2007aw}). Even relatively small contributions to such an energy density is visible: At early times it leads to spectral distortions in the CMB, and at late times stringent bounds come from the non-observation of high energy particles.

Dark matter decaying to other particles in the dark sector is much less constrained, and several qualitatively different scenarios can arise. 
If dark matter decays into relativistic particles (dark radiation), it is still true that at most a small fraction of all dark matter can have decayed before the present. Otherwise structure formation would be strongly suppressed and this leads to conflict with numerous types of observations.

A different scenario which has been studied in some detail is the case where a cold dark matter particle decays to a slightly lighter dark matter particle, such that the daughter particle is highly non-relativistic. This would add a small velocity dispersion to the resulting dark matter component and suppress small scale structure growth (see e.g.\ \cite{SanchezSalcedo:2003pb,Peter:2010au,Bell:2010qt,Wang:2013rha,Cheng:2015dga,Vattis:2019efj}). 

In this paper we will look at the simplest possible case where some fraction of cold dark matter is unstable and decay to dark radiation (see \cite{Takahashi:2003iu,Audren:2014bca,Enqvist:2015ara,Poulin:2016nat,Bringmann:2018jpr}).
Most studies of dark matter decay have been done purely in linear perturbation theory, which is only valid on relatively large scales mainly probed by CMB observations.
Some studies have looked into dark matter decay on smaller scales, using $N$-body simulations (see e.g.\ \cite{suto1988,Enqvist:2015ara}). In these cases the dark matter particle mass in the simulation is typically downscaled by a factor corresponding exactly to how the background density of dark matter decreases due to decay.

However, this approach neglects several effects which are potentially noticeable: 1) It ignores the change to the background expansion rate caused by the conversion of non-relativistic matter into radiation. 2) It ignores perturbations in the dark radiation component created through dark matter decay. 3) It neglects the gravitational correction to the lifetime of the dark matter particles.
Point 1) can easily be incorporated by solving the correct background equations to find the scale factor $a(t)$. However, the two other points are non-trivial and require additional information to be fed to the $N$-body code.

We will approach point 2) by solving the full Einstein-Boltzmann equations in linear perturbation theory and subsequently add perturbations from all species which are not dark matter or baryons to the simulation using the prescription outlined in \cite{Tram:2018znz}.
Point 3) can be treated using the same formalism because the potential governing the correction to the decay rate can be realised in the simulation.
Using the formalism from \cite{Tram:2018znz} also automatically takes into account general relativistic corrections to the equations of motion of particles in the simulation and the whole framework is thus fully compatible with general relativity.

In section 2 we outline the formalism needed and present our numerical implementation. Section 3 contains a description of the main numerical results, and finally section 4 contains a discussion and conclusion.

%%%%%%%%%%%%%%%%%%%%%%%%%%%%%%%%%%%%%%%%%%%%%%%%%%%%%%%%%%%%%%%%%%%%%%%%
\section{Method and implementation}

\subsection{Linear theory}

At the background level, decaying dark matter (`dcdm') and dark radiation (`dr') distinguish themselves by having a source term in their continuity equations;
\begin{align}
	\dot{\bar{\rho}}_{\text{dcdm}} &= -3\mathcal{H}\bar{\rho}_{\text{dcdm}} - a\Gamma_{\text{dcdm}}\bar{\rho}_{\text{dcdm}} \,, \label{eq:homo_continuity_dcdm}  \\
	\dot{\bar{\rho}}_{\text{dr}} &= -4\mathcal{H}\bar{\rho}_{\text{dr\phantom{dm}}} + a\Gamma_{\text{dcdm}}\bar{\rho}_{\text{dcdm}}\,, \label{eq:homo_continuity_dr} 
\end{align}
which serves to pump energy from the decaying dark matter component into the dark radiation component. Here a dot denotes differentiation with respect to conformal time $\tau$ (defined through $\mathrm{d}\tau = \mathrm{d}t/a(t)$, $t$ being cosmic time) and $\mathcal{H}\equiv\dot{a}/a$ is the conformal Hubble parameter with $a$ the cosmic scale factor. Lastly, $\Gamma_{\text{dcdm}}$ is the constant decay rate of decaying dark matter, defined with respect to proper time.

At the linear perturbation level, the full general relativistic continuity and Euler equations for decaying dark matter in $N$-body gauge (superscript `Nb') may be written as \cite{Fidler:2017ebh}
\begin{align}
	\dot{\delta}_{\text{dcdm}}^{\text{Nb}} + \nabla\cdot \tilde{\vec{v}}_{\text{dcdm}}^{\text{Nb}} &= -\frac{1}{3}\frac{\Gamma_{\text{dcdm}}}{H} \dot{H}_{\text{T}}^{\text{Nb}} \,, \label{eq:continuity_dcdm} \\
	(\partial_\tau + \mathcal{H})\tilde{\vec{v}}_{\text{dcdm}}^{\text{Nb}} &= -\nabla(\phi - \gamma^{\text{Nb}})\,, \label{eq:euler_dcdm}
\end{align}
where $H\equiv \dot{a}/a^2$ is the Hubble parameter and $H_{\text{T}}^{\text{Nb}}$ is the trace-free component of the spatial part of the metric in $N$-body gauge (see e.g.\ \cite{Adamek:2017grt}). The fluid velocity $\tilde{\vec{v}}_{\text{dcdm}}$ would normally be written without the tilde. The reason for this naming choice will become apparent at the end of this subsection.
Note that both continuity equations \eqref{eq:homo_continuity_dcdm} and \eqref{eq:continuity_dcdm} revert back to their usual form for stable cold dark matter simply be setting $\Gamma_{\text{dcdm}}=0$, and that the Euler equation \eqref{eq:euler_dcdm} is identical to that of stable cold dark matter (see e.g.\ \cite{Fidler:2017ebh,Tram:2018znz}).

The gauge-invariant gravitational potential $\phi$ satisfies a Newtonian Poisson equation in $N$-body gauge with contributions from all species,
\begin{equation}
	\nabla^2\phi = 4\pi Ga^2\sum_\alpha \delta\rho_\alpha^{\text{Nb}}\,. \label{eq:Poisson}
\end{equation}
In the case of a standard $\Lambda\text{CDM}$ cosmology augmented with decaying dark matter and dark radiation, we have $\alpha\in\{\text{b}, \text{cdm}, \text{dcdm}, \gamma, \nu, \text{dr}\}$, where we allow for `cdm' (stable cold dark matter) and `dcdm' to coexist as separate species. We obtain all linear transfer functions from \CLASS{}, which can operate and produce output in either synchronous `s' or conformal Newtonian `N' gauge. To convert to $N$-body gauge we use
\begin{align}
	\delta\rho_\alpha^{\text{Nb}} &= \delta\rho_\alpha^{\text{s/N}} - \dot{\bar{\rho}}_\alpha \frac{\theta^{\text{s/N}}_{\text{tot}}}{k^2} \notag \\
	&= \delta\rho_\alpha^{\text{s/N}} + \bigl[3\mathcal{H}(1 + w_\alpha) + a\Gamma_\alpha\bigr]\bar{\rho}_\alpha\frac{\theta^{\text{s/N}}_{\text{tot}}}{k^2}\,, \label{eq:deltarho_gauge_transform}
\end{align}
where the equation of state $w_\alpha \equiv \bar{p}_\alpha/\bar{\rho}_\alpha$, $\theta_{\text{tot}}$ is the total velocity divergence of all species and the appearance of $\Gamma_\alpha$ matches \eqref{eq:homo_continuity_dcdm}. Explicitly setting $\Gamma_\alpha = 0$ for stable species allows us to use \eqref{eq:deltarho_gauge_transform} for these as well. 
We can even generalise \eqref{eq:deltarho_gauge_transform} to be applicable to stable species which are the decay products of other species (here only dark radiation) by  generalising $\Gamma_\alpha$ further as
\begin{equation}
	\Gamma_{\text{dr}} \equiv -\frac{\bar{\rho}_{\text{dcdm}}}{\bar{\rho}_{\text{dr}}}\Gamma_{\text{dcdm}}\,.
\end{equation}
However, note that the ``decay'' rate for decay product species is then not an actual physical quantity but an auxiliary quantity (for example, it is negative and time dependent).

The correction potential $\gamma^{\text{Nb}}$ in \eqref{eq:euler_dcdm} originates in perturbed non-dust components such as relativistic species. We refer the reader to \cite{Tram:2018znz} for details on how to compute this quantity. The only difference here is that dark radiation contribute to the total pressure perturbation, total shear and the time derivative of the total background pressure, all of which $\gamma^{\text{Nb}}$ depends upon. These contributions are trivially included, since dark radiation behaves just like any ultra-relativistic species, with the exception of the source term in \eqref{eq:homo_continuity_dr}. The time derivative of the background pressure $\bar{p}_{\text{dr}}$ then similarly gets a source term;
\begin{equation}
	\dot{\bar{p}}_{\text{dr}} = -\frac{4}{3}\mathcal{H} \bar{\rho}_{\text{dr}} + \frac{1}{3}a\Gamma_{\text{dcdm}}\bar{\rho}_{\text{dcdm}} \,.
\end{equation}

\begin{figure}[t]
\begin{center}
\includegraphics[width=0.85\textwidth]{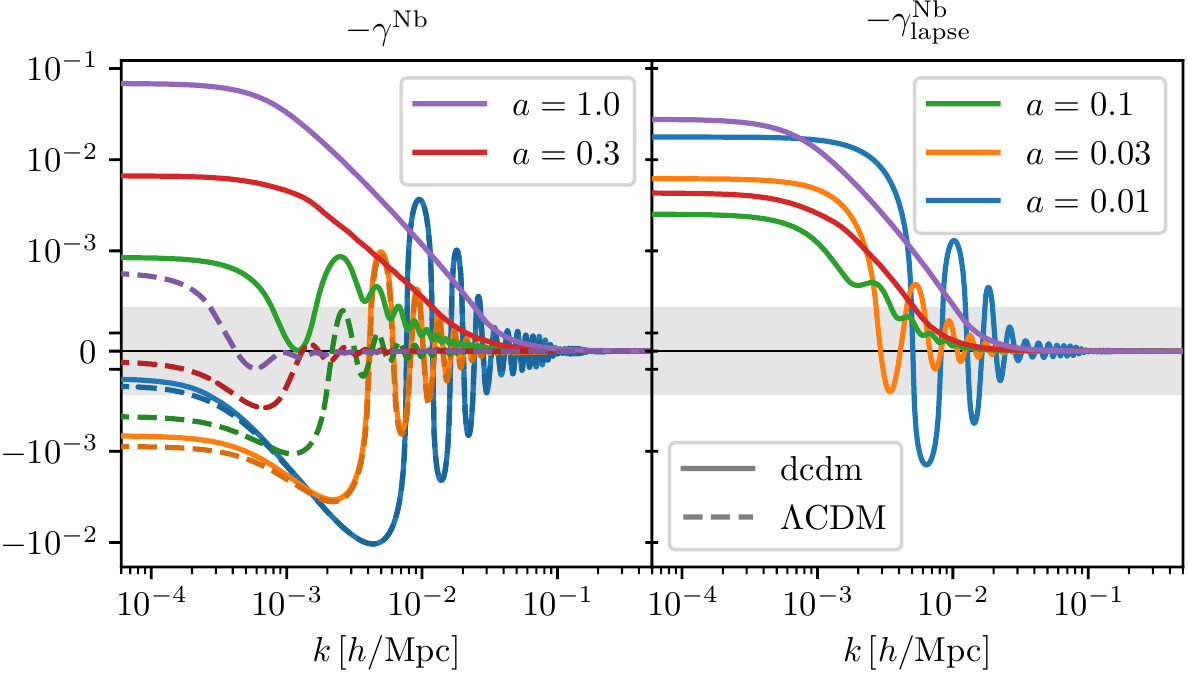}
\end{center}
\caption{Fourier space plots of the GR correction potentials $\gamma^{\text{Nb}}$ and $\gamma^{\text{Nb}}_{\text{lapse}}$ at different times. Full lines show the case of $\Gamma_{\text{dcdm}}=\SI{10}{km.s^{-1}.Mpc^{-1}}$ and where $f_{\text{dcdm}}=0.3$ of all dark matter is of the decaying kind (see subsection~\ref{subsec:sim_params} for a precise definition of $f_{\text{dcdm}}$). Dashed lines show the case of $\Lambda$CDM. In both cases the remaining cosmological parameters are as specified in table~\ref{table:class_parameters}. The grey band indicates the region where the vertical axis scales linearly.
\label{fig:lapse}}
\end{figure}

The decay rate $\Gamma_{\text{dcdm}}$ for a given decaying dark matter particle is measured with respect to the proper time for that particle. When writing the equations of motion using a globally defined clock (here $\tau$), a correction term is needed, which is exactly the origin of the source term in the continuity equation \eqref{eq:continuity_dcdm}. We wish to reuse the numerical machinery of \cite{Tram:2018znz} for applying the Euler equation correction $\gamma^{\text{Nb}}$ to also include this source term. By defining $\vec{v}_{\text{dcdm}} \equiv \tilde{\vec{v}}_{\text{dcdm}} + \Gamma_{\text{dcdm}}/(3H)\nabla^{-1}\dot{H}_{\text{T}}^{\text{Nb}}$ we can write the equations of motion for decaying dark matter \eqref{eq:continuity_dcdm} and \eqref{eq:euler_dcdm} as
\begin{align}
	\dot{\delta}_{\text{dcdm}}^{\text{Nb}} + \nabla\cdot \vec{v}_{\text{dcdm}}^{\text{Nb}} &= 0 \,, \label{eq:continuity_dcdm_newv} \\
	(\partial_\tau + \mathcal{H})\vec{v}_{\text{dcdm}}^{\text{Nb}} &= -\nabla\biggl(\phi - \gamma^{\text{Nb}} - \frac{\Gamma_{\text{dcdm}}}{H}\gamma_{\text{lapse}}^{\text{Nb}}\biggr)\,, \label{eq:euler_dcdm_newv}
\end{align}
where the lapse potential $\gamma_{\text{lapse}}$ is a new general relativistic correction potential, given in $N$-body gauge as
\begin{equation}
	\gamma_{\text{lapse}}^{\text{Nb}} = \frac{1}{3}\nabla^{-2}\biggl[\biggl(2\mathcal{H} - \frac{\dot{\mathcal{H}}}{\mathcal{H}}\biggr)\dot{H}_{\text{T}} + \ddot{H}_{\text{T}} \biggr]\,. \label{eq:gamma_lapse}
\end{equation}
Leaving $\Gamma_{\text{dcdm}}/H$ out of the definition of $\gamma_{\text{lapse}}$ decouples this potential from anything component-specific, and the equations \eqref{eq:continuity_dcdm_newv} and \eqref{eq:euler_dcdm_newv} may then be taken as the general equations of motion for any matter component by a trivial substitution of species subscripts, remembering that we have $\Gamma_{\text{b}}=\Gamma_{\text{cdm}}=0$.

In Fig.~\ref{fig:lapse} we plot both $\gamma^{\text{Nb}}$ and $\gamma^{\text{Nb}}_{\text{lapse}}$ for a sample dcdm cosmology, from which we see that both exhibit complex oscillatory behaviour. The figure also shows $\gamma^{\text{Nb}}$ in the case of $\Lambda$CDM, from which we see that $\gamma^{\text{Nb}}$ is very similar at early times but orders of magnitudes greater at late times in the case of dcdm, caused by the large generation of dark radiation.

For the case shown in Fig.~\ref{fig:lapse} we see that the two correction potentials are of comparable size, which happens to be the case generally. However, the factor $\Gamma_{\text{dcdm}}/H$ in \eqref{eq:euler_dcdm_newv} is never much greater than unity for reasonable values\footnote{See subsection~\ref{subsec:sim_params} for experimental bounds on $\Gamma_{\text{dcdm}}$.} of $\Gamma_{\text{dcdm}}$, and decreases rapidly as we go back in time. While both correction potentials are only relevant at large scales, as expected, $\gamma^{\text{Nb}}_{\text{lapse}}$ is then furthermore only relevant at late times.

\subsection{Non-linear implementation}
As in \cite{Tram:2018znz} we write the potential $\phi-\gamma^{\text{Nb}}$ of \eqref{eq:euler_dcdm_newv} in terms of a contribution $\phi_{\text{sim}}$ from the matter particles in the simulation and a contribution $\phi_{\text{GR}}$ from all other species (typically photons and neutrinos but now also dark radiation) as well as the GR effects supplied by $\gamma^{\text{Nb}}$:
\begin{equation}
	\phi - \gamma^{\text{Nb}} = \phi_{\text{sim}} + \phi_{\text{GR}}
\end{equation}
with
\begin{equation}
\nabla^2\phi_{\text{GR}} = 4\pi Ga^2(\delta\rho_{\gamma}^{\text{Nb}} + \delta\rho_{\nu}^{\text{Nb}} + \delta\rho_{\text{dr}}^{\text{Nb}})-\nabla^2\gamma^{\text{Nb}} \,. \label{eq:phi_GR} 
\end{equation}
With this, the Euler equation \eqref{eq:euler_dcdm_newv} may be written
\begin{equation}
    (\partial_\tau + \mathcal{H})\vec{v}_{\text{m}}^{\text{Nb}} = -\nabla\phi_{\text{sim}} - \nabla\biggl(\phi_{\text{GR}} - \frac{\Gamma_{\text{m}}}{H} \gamma_{\text{lapse}}^{\text{Nb}}\biggr)\,, \label{eq:euler_final}
\end{equation}
where subscript $\text{m}\in\{\text{b}, \text{cdm}, \text{dcdm}\}$ indicates any matter species. The first term on the right in \eqref{eq:euler_final} is then the usual non-linear gravity due to matter computed via standard $N$-body techniques, while the last term is the linear correction, itself comprised of a general term $\phi_{\text{GR}}$ and a term unique to decaying matter $\propto\gamma_{\text{lapse}}^{\text{Nb}}$.

We have fully implemented the physics of decaying cold dark matter laid out in the previous subsection into the publicly available \CONCEPT{} $N$-body code \cite{Dakin:2017idt}, using the non-linear/linear splitting of gravity \eqref{eq:euler_final}. As \CONCEPT{} inherits the cosmological background directly from \CLASS{}, the correct $a(t)$ and hence overall growth of structure is trivially obtained also in the presence of decaying dark matter. Similarly, the linear gravitational effects on matter from dark radiation perturbations are obtained simply by including their contributions (consisting of $\delta\rho_{\text{dr}}^{\text{Nb}}$ and contributions to $\gamma^{\text{Nb}}$) to $\phi_{\text{GR}}$ \eqref{eq:phi_GR}, as described for photons and neutrinos in \cite{Tram:2018znz}. Using \eqref{eq:gamma_lapse} with $\dot{H}_{\text{T}}$ from \CLASS{}, we have implemented $\gamma^{\text{Nb}}_{\text{lapse}}$ in a fashion similar to that of $\gamma^{\text{Nb}}$, as described in \cite{Tram:2018znz}. What is left is the implementation of mass reduction of the $N$-body particles in the non-linear simulation.

\subsubsection*{Mass reduction}
The particles of an $N$-body simulation are fully specified by their comoving coordinates $\vec{x}_{\text{m},i}$ and canonical momenta $\vec{q}_{\text{m},i}$, where $i=1,\dots,N$ labels the individual particles. To apply the Eulerian equations of motion for matter to the particles, the equations must first be expressed in Lagrangian form. For the general equations \eqref{eq:continuity_dcdm_newv} and \eqref{eq:euler_final} including general relativistic effects and allowing for decaying cold dark matter, we have 
\begin{align}
	\partial_t \vec{x}_{\text{m},i} &= \frac{\vec{q}_{\text{m},i}}{a^2m_{\text{m}}} \,,  \label{eq:xdot} \\
	\partial_t \vec{q}_{\text{m},i} &= -\Gamma_{\text{m}}\vec{q}_{\text{m},i} -m_{\text{m}}\nabla\phi_{\text{sim}}\big\rvert_{\vec{x}=\vec{x}_{\text{m},i}} - m_{\text{m}}\nabla\biggl(\phi_{\text{GR}} - \frac{\Gamma_{\text{m}}}{H}\gamma^{\text{Nb}}_{\text{lapse}}\biggr)\bigg\rvert_{\vec{x}=\vec{x}_{\text{m},i}} \,, \label{eq:qdot}
\end{align}
where we have switched to cosmic time $t$ as this is the time variable used within \CONCEPT{} (and most other $N$-body codes) and the mass $m_{\text{m}}$ is the same for all particles belonging to a given species $\text{m}$. Each $N$-body particle is thought of as being constituted by a near-infinite number of elementary particles, and so the $N$-body particle mass falls off as $\partial_t m_{\text{m}}=-\Gamma_{\text{m}} m_{\text{m}} \Rightarrow m_{\text{m}}\propto \text{e}^{-\Gamma_{\text{m}}t}$. As this drop in mass does not change the velocity $\vec{q}_{\text{m},i}/m_{\text{m}}$, the momentum must instead decrease as given by the first term in \eqref{eq:qdot}.

The numerical implementation of the Lagrangian equations of motion \eqref{eq:xdot} and \eqref{eq:qdot} are often written in terms of drift $D(\Delta t)$ and kick $K(\Delta t)$ operators (see e.g.\ \cite{Springel:2005mi}), which in our case may be written
\begin{align}
    D(\Delta t) \{\vec{x}_{\text{m},i}(t)\} &= \biggl\{ \vec{x}_{\text{m},i}(t) + \frac{\vec{q}_{\text{m},i}(t)}{m_{\text{m}}(t)}\int_t^{t+\Delta t} \frac{\mathrm{d}t'}{a^2(t')}\biggr\}\,, \label{eq:drift_op} \\
    K(\Delta t) \{\vec{q}_{\text{m},i}(t)\} &= \biggl\{ \text{e}^{-\Gamma_{\text{m}}\Delta t}\vec{q}_{\text{m},i}(t) - \nabla\phi_{\text{sim}}(t)\big\rvert_{\vec{x}=\vec{x}_{\text{m},i}}a(t)\int_t^{t+\Delta t} \frac{m_{\text{m}}(t')}{a(t')}\,\text{d}t' \notag \\  &\qquad - \nabla\biggl( \int_t^{t+\Delta t}m_{\text{m}}(t')\phi_{\text{GR}}(t') \,\text{d}t' - \Gamma_{\text{m}}\int_t^{t+\Delta t}\frac{m_{\text{m}}(t')}{H(t')} \gamma_{\text{lapse}}^{\text{Nb}}(t') \,\text{d}t'\biggr)\bigg\rvert_{\vec{x}=\vec{x}_{\text{m},i}}\biggr\} \,, \label{eq:kick_op}
\end{align}
where $\{\vec{x}_{\text{m},i}(t+\Delta t)\} \approx D(\Delta t) \{\vec{x}_{\text{m},i}(t)\}$ and $\{\vec{q}_{\text{m},i}(t+\Delta t)\} \approx K(\Delta t) \{\vec{q}_{\text{m},i}(t)\}$ are used to numerically integrate the particle system forwards in time by an amount $\Delta t$. In \CONCEPT{} time integration is performed using the leapfrog method with a global time step size $\Delta t$.

The integrals in \eqref{eq:drift_op} and \eqref{eq:kick_op} are performed in order to obtain average values over the time step of functions that are continuously defined. In \eqref{eq:drift_op}, for example, we could replace the integral with e.g.\ $\Delta t/a^2(t)$, at the cost of numerical precision. For the exact solution, we should include the whole term under the integral, i.e.\ we should additionally include the momentum $\vec{q}_{\text{m},i}$ and mass $m_{\text{m}}$. However, the momentum is only integrable in discrete steps using the kick operator $K$, and so numerically we are unable to include $\vec{q}_{\text{m},i}$ under the integral in \eqref{eq:drift_op}. Contrary, we \emph{could} include $m_{\text{m}}(t')$ under the integral as it is known at all times. However, as the time dependence $m_{\text{m}}\propto \text{e}^{-\Gamma_{\text{m}}t}$ due to decay is shared\footnote{Of course the presence of external forces ($\phi_{\text{sim}}$, $\phi_{\text{GR}}$, $\gamma_{\text{lapse}}^{\text{Nb}}$) adds to the full time dependence of $\vec{q}_{\text{m},i}$.} by $\vec{q}_{\text{m},i}$ and this is outside the integral, we need $m_{\text{m}}$ outside the integral as well, and evaluated at the same time $t$.

As we have the linear correction potentials $\phi_{\text{GR}}$ and $\gamma_{\text{lapse}}^{\text{Nb}}$ tabulated on an $(a, k)$ grid from \CLASS{}, we can include these under the integrals in \eqref{eq:kick_op} by performing the integral for each $k$ mode prior to the real-space realisation.

We cannot include the non-linear potential $\phi_{\text{sim}}$ under the integral in \eqref{eq:kick_op}, as this is only evaluated at the discrete time steps. As this potential arise solely from the matter species present we have $\phi_{\text{sim}}\appropto a^2(\bar{\rho}_{\text{b}}+\bar{\rho}_{\text{cdm}}+\bar{\rho}_{\text{dcdm}})\appropto a^{-1}$, and so instead we integrate over this temporal dependency.

As described in \cite{Dakin:2017idt}, \CONCEPT{} makes heavy use of the parametrisation $\varrho_\alpha \equiv a^{3(1 + \mathscr{w}_\alpha)}\rho_\alpha$, where $\mathscr{w}_\alpha(a)$ is called the effective equation of state. By definition $\bar{\varrho}_\alpha$ is constant in time and equal to $\bar{\rho}_\alpha(a=1)$, from which we see that
\begin{equation}
    \mathscr{w}_\alpha = \frac{\ln \bar{\rho}_\alpha(a=1)/\bar{\rho}_\alpha}{3\ln a} - 1\,. \label{eq:eos_eff}
\end{equation}
In the case of stable matter for which $\bar{\rho}_{\text{m}}\propto a^{-3}$, $\mathscr{w}_{\text{m}}$ reduces to the usual equation of state $\mathscr{w}_{\text{m}}=w_{\text{m}}=0$. For decaying matter where the time dependence of $\bar{\rho}_{\text{m}}$ is less trivial, $\mathscr{w}_{\text{m}}$ becomes non-zero and time dependent. With $\bar{\rho}_{\text{m}}(a)$ tabulated by \CLASS{} we can now compute $\mathscr{w}_{\text{m}}(a)$ using \eqref{eq:eos_eff}. We then have the general time dependence $\bar{\rho}_{\text{m}}\propto a^{-3(1+\mathscr{w}_{\text{m}})}$, of which $a^{-3}$ is due to the Hubble expansion. Thus, the mass of each $N$-body particle decays away as $m_{\text{m}}(a) = m_{\text{m},0} a^{-3\mathscr{w}_{\text{m}}}$ with $m_{\text{m},0} \equiv m_{\text{m}}(a=1)$.

Instead of discretely reducing the particle mass between each time step, we introduce the continuous scaling of $m_{\text{m}}(a)$ into the drift \eqref{eq:drift_op} and kick \eqref{eq:kick_op} operator:
\begin{align}
    D(\Delta t) \{\vec{x}_{\text{m},i}(t)\} &= \biggl\{ \vec{x}_{\text{m},i}(t) + \frac{\vec{q}_{\text{m},i}(t)}{m_{\text{m},0}} \int_t^{t+\Delta t}  a(t')^{3\mathscr{w}_{\text{m}}(t') - 2} \,\mathrm{d}t'\biggr\}\,, \label{eq:drift_op_final} \\
    K(\Delta t) \{\vec{q}_{\text{m},i}(t)\} &= \biggl\{ \frac{a(t)^{3\mathscr{w}_{\text{m}}(t)}}{a(t+\Delta t)^{3\mathscr{w}_{\text{m}}(t+\Delta t)}} \vec{q}_{\text{m},i}(t) \notag \\
    &\qquad - m_{\text{m},0}\nabla\phi_{\text{sim}}(t)\big\rvert_{\vec{x}=\vec{x}_{\text{m},i}}a(t)^{3\mathscr{w}_{\text{M}}(t) + 1}\int_t^{t+\Delta t} a(t')^{-3\mathscr{w}_{\text{m}}(t') -3\mathscr{w}_{\text{M}}(t') - 1} \,\text{d}t' \notag \\
    &\qquad - m_{\text{m},0}\nabla\biggl( \int_t^{t+\Delta t}a(t')^{-3\mathscr{w}_{\text{m}}(t')}\phi_{\text{GR}}(t') \,\text{d}t' \notag \\
    &\qquad\qquad\qquad\quad - \int_t^{t+\Delta t}\frac{\Gamma_{\text{m}}(t')}{H(t')}a(t')^{-3\mathscr{w}_{\text{m}}(t')} \gamma_{\text{lapse}}^{\text{Nb}}(t') \,\text{d}t'\biggr)\bigg\rvert_{\vec{x}=\vec{x}_{\text{m},i}}\biggr\} \,, \label{eq:kick_op_final}
\end{align}
where we now also take the decay into account regarding the scaling of $\phi_{\text{sim}}$ using $\phi_{\text{sim}}\appropto a^2(\bar{\rho}_{\text{b}}+\bar{\rho}_{\text{cdm}}+\bar{\rho}_{\text{dcdm}})\propto a^{3\mathscr{w}_{\text{b}+\text{cdm}+\text{dcdm}}-1} \equiv a^{3\mathscr{w}_{\text{M}}-1}$ obtained from \eqref{eq:eos_eff} using
\begin{equation}
\bar{\rho}_{\text{M}} \equiv \bar{\rho}_{\text{b}+\text{cdm}+\text{dcdm}} \equiv \bar{\rho}_{\text{b}} + \bar{\rho}_{\text{cdm}} + \bar{\rho}_{\text{dcdm}}\,.
\end{equation}
For simulations without decaying matter, $\mathscr{w}_{\text{M}} = 0$ and so the explicit scaling of $\phi_{\text{sim}}$ reduces back to $a^{-1}$. Equation \eqref{eq:kick_op_final} is additionally written to allow for a time dependent $\Gamma_{\text{m}}(t)$, in which case $m_{\text{m}}\propto \text{e}^{-\Gamma_{\text{m}}t}$ no longer holds while $m_{\text{m}}\propto a^{-3\mathscr{w}_{\text{m}}}$ still does. We shall see how this additional generalisation becomes important when we collect together the different matter species into a single component.

\subsubsection*{Combining matter species}

It is customary to combine baryons and cold dark matter into a single $N$-body component, instead of evolving these as two separate sets of particles. 
Neglecting baryonic (gas) physics\footnote{This is customary in most cosmological $N$-body simulations.}, this can be done trivially because the 
equations of motions for the two species are identical, so that they differ only in their initial conditions.

One might wish to also add decaying cold dark matter to this combined matter component. From \eqref{eq:homo_continuity_dcdm}, \eqref{eq:continuity_dcdm_newv} and \eqref{eq:euler_dcdm_newv}, one can easily show that
\begin{align}
    \dot{\delta}_{\text{M}}^{\text{Nb}} + \nabla\cdot\vec{v}_{\text{M}}^\text{Nb} &= a\Gamma_{\text{M}}\bigl(\delta_{\text{M}}^\text{Nb} - \delta_{\text{dcdm}}^\text{Nb}\bigr) \,, \label{eq:continuity_combined} \\
    (\partial_\tau + \mathcal{H})\vec{v}_{\text{M}}^\text{Nb} &= -\nabla\biggl(\phi - \gamma^\text{Nb} - \frac{\Gamma_{\text{M}}}{H}\gamma_{\text{lapse}}^{\text{Nb}}\biggr) + a\Gamma_{\text{M}}\bigl(\vec{v}_{\text{M}}^\text{Nb} - \vec{v}_{\text{dcdm}}^\text{Nb}\bigr)\,, \label{eq:euler_combined}
\end{align}
where
\begin{align}
    \delta_{\text{M}} &\equiv \delta_{\text{b}+\text{cdm}+\text{dcdm}}  = \frac{1}{\bar{\rho}_{\text{M}}} \,\,\,\,\,\,\,\,\sum_{\mathclap{\,\,\,\,\,\,\,\,\,\,\,\,\,\,\,\,\,\,\text{m}\in\{\text{b},\, \text{cdm},\, \text{dcdm}\}}} \delta_{\text{m}}\bar{\rho}_{\text{m}}\,,
 \\
    \vec{v}_{\text{M}} &\equiv \vec{v}_{\text{b}+\text{cdm}+\text{dcdm}} = \frac{1}{\bar{\rho}_{\text{M}}} \,\,\,\,\,\,\,\,\sum_{\mathclap{\,\,\,\,\,\,\,\,\,\,\,\,\,\,\,\,\,\,\text{m}\in\{\text{b},\, \text{cdm},\, \text{dcdm}\}}} \vec{v}_{\text{m}}\bar{\rho}_{\text{m}} \,, \\
    \Gamma_{\text{M}} &\equiv \Gamma_{\text{b}+\text{cdm}+\text{dcdm}} = \frac{1}{\bar{\rho}_{\text{M}}} \,\,\,\,\,\,\,\,\sum_{\mathclap{\,\,\,\,\,\,\,\,\,\,\,\,\,\,\,\,\,\,\text{m}\in\{\text{b},\, \text{cdm},\, \text{dcdm}\}}} \Gamma_{\text{m}}\bar{\rho}_{\text{m}}\,\,\,\,\,  = \frac{\bar{\rho}_{\text{dcdm}}}{\bar{\rho}_{\text{M}}} \Gamma_{\text{dcdm}} \,. \label{eq:Gamma_combined}
\end{align}

\addtocounter{footnote}{-1}  % For the footnote in figure fig:delta_theta
\begin{figure}[t]
\begin{center}
\includegraphics[width=0.9\textwidth]{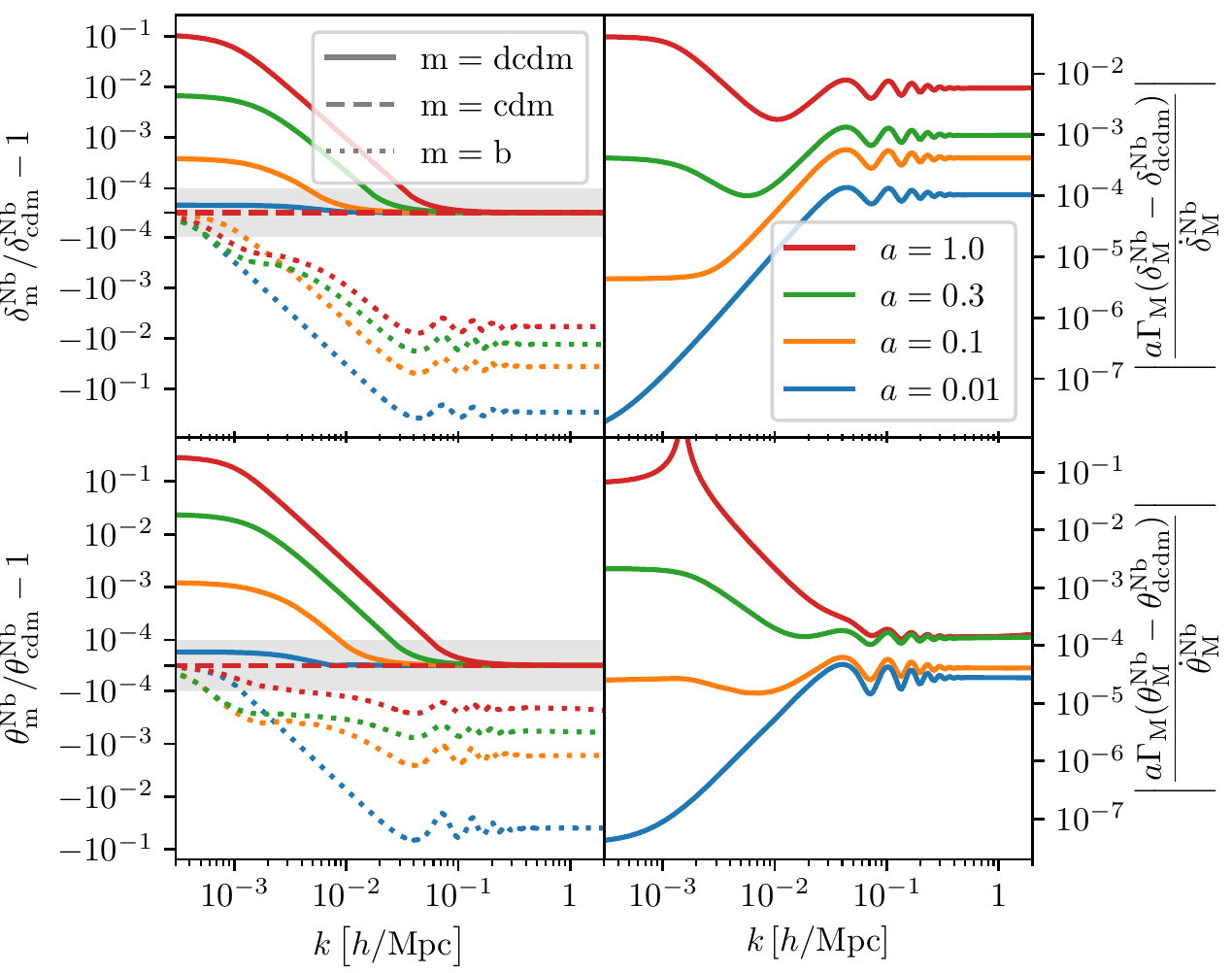}
\end{center}
\caption[bla]{Linear evolution of $\delta_{\text{m}}^{\text{Nb}}$ and $\theta_{\text{m}}^{\text{Nb}}\equiv \nabla\cdot\vec{v}_{\text{m}}^{\text{Nb}}$ in a cosmology where all of the cold dark matter is decaying, with $\Gamma_{\text{dcdm}}=\SI{100}{km.s^{-1}.Mpc^{-1}}$. Other cosmological parameters are as given in table~\ref{table:class_parameters}. The left panels show the evolution of decaying cold dark matter and baryons relative to that of stable cold dark matter\protect\footnotemark. Here the grey bands indicate regions where the vertical axes scale linearly. The right panels show the size of the additional terms on the right-hand side of the equations of motion for the combined matter component \eqref{eq:continuity_combined} and \eqref{eq:euler_combined}, compared to the entire time derivatives $\dot{\delta}_{\text{M}}^{\text{Nb}}$ and $\dot{\theta}_{\text{M}}^{\text{Nb}}$.
\label{fig:delta_theta}}
\end{figure}
% We have to write the footnote in figure fig:delta_theta like this
\footnotetext{To make sense of $\delta_{\text{cdm}}$ and $\theta_{\text{cdm}}$ in the cosmology of Fig.~\ref{fig:delta_theta} we might imagine adding a negligible amount of stable cold dark matter, as otherwise all dark matter is of the decaying kind. As $\delta_{\text{cdm}}$ and $\theta_{\text{cdm}}$ do not depend strongly on the amount of stable cold dark matter but only on the amount of total matter, this is perfectly fine.}

\noindent We see that combining all three matter species introduces the additional terms $\propto \bigl(\delta_{\text{M}}^{\text{Nb}} - \delta_{\text{dcdm}}^{\text{Nb}}\bigr)$ and $\propto \bigl(\vec{v}_{\text{M}}^{\text{Nb}} - \vec{v}_{\text{dcdm}}^{\text{Nb}}\bigr)$ in the continuity \eqref{eq:continuity_combined} and Euler \eqref{eq:euler_combined} equation, respectively, compared to their single-species versions \eqref{eq:continuity_dcdm_newv} and \eqref{eq:euler_dcdm_newv}. The combined system then cannot be solved exactly without separately solving the decaying dark matter species. This is entirely due to the decay source term in \eqref{eq:homo_continuity_dcdm}, as the combined $\Gamma_{\text{M}}$ \eqref{eq:Gamma_combined} do allow the lapse potential to act on the combined system \eqref{eq:euler_combined} in exactly the same manner as in \eqref{eq:euler_dcdm_newv}. Note that $\Gamma_{\text{M}}$ changes with time.

We might argue that $\bigl(\delta_{\text{M}}^{\text{Nb}} - \delta_{\text{dcdm}}^{\text{Nb}}\bigr)$ and $\bigl|\vec{v}_{\text{M}}^{\text{Nb}} - \vec{v}_{\text{dcdm}}^{\text{Nb}}\bigr|$ ought to be small, as we expect the difference between decaying and stable matter to be large only at the background level. Furthermore, these terms are multiplied by $\Gamma_{\text{M}}\propto\Gamma_{\text{dcdm}}$ in \eqref{eq:continuity_combined} and \eqref{eq:euler_combined} and so vanish completely in the limit $\Gamma_{\text{dcdm}} \rightarrow 0$, as expected. In Fig.~\ref{fig:delta_theta} we plot the extreme case of having all cold dark matter be of the decaying kind, while at the same time having a large value of $\Gamma_{\text{dcdm}}=\SI{100}{km.s^{-1}.Mpc^{-1}}$. The right panels show that indeed these new terms in the continuity \eqref{eq:continuity_combined} and Euler \eqref{eq:euler_combined} equation for the combined matter component are always subdominant (the divergence on the lower right panel at $a=1$, $k\sim 1.5\times 10^{-3}\, h/\text{Mpc}$ is caused by a sign change in $\dot{\theta}_{\text{M}}^{\text{Nb}}$).

\begin{figure}[t]
\begin{center}
\includegraphics[width=0.85\textwidth]{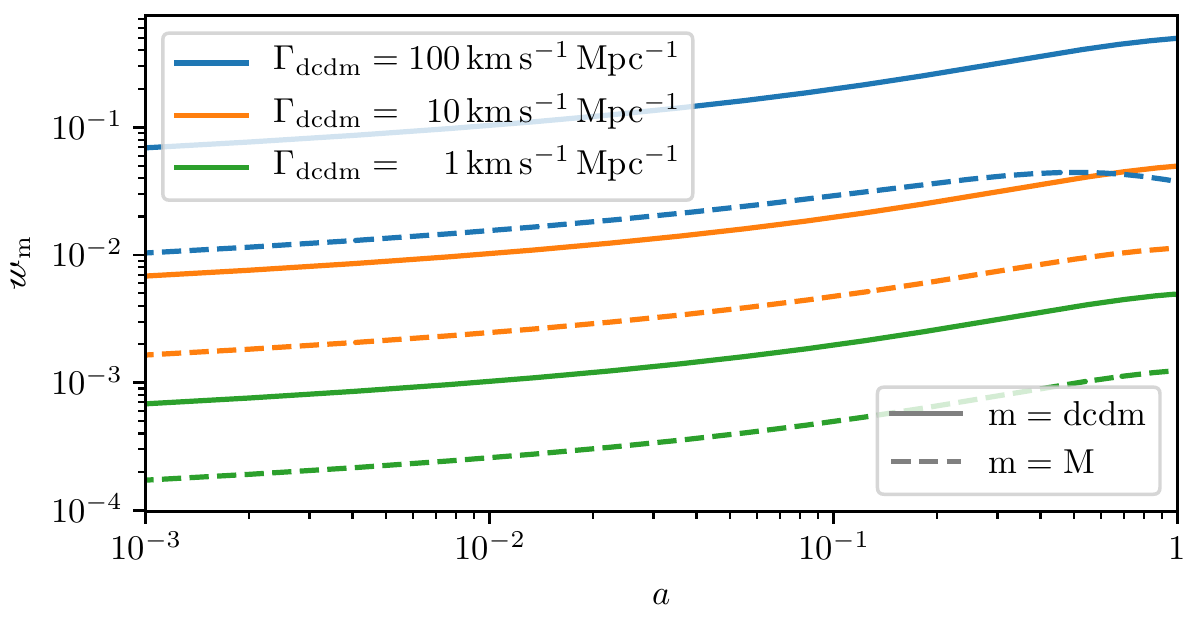}
\end{center}
\caption{Evolution of the effective equation of state $\mathscr{w}_{\text{m}}(a)$ for decaying cold dark matter (solid lines) and total matter (dashed lines) in the cosmology specified by table~\ref{table:class_parameters}, for three different values of $\Gamma_{\text{dcdm}}$. In the case of total matter $m=\text{M}\equiv\text{b}+\text{cdm}+\text{dcdm}$, $\mathscr{w}_{\text{m}}(a)$ lies somewhere between $\mathscr{w}_{\text{b}}=\mathscr{w}_{\text{cdm}}=0$ and $\mathscr{w}_{\text{dcdm}}(a)$ dependent on the relative abundance of the different matter species. For the case shown, $f_{\text{dcdm}}=0.3$ of all dark matter is of the decaying kind (see subsection~\ref{subsec:sim_params} for a precise definition of $f_{\text{dcdm}}$).
\label{fig:w_eff}}
\end{figure}

Now looking at the left panels of Fig.~\ref{fig:delta_theta}, we see that even in the extreme case shown the difference between decaying cold dark matter and stable cold dark matter is comparable to the difference between baryons and stable cold dark matter. While baryons have less structure than stable cold dark matter at sub-horizon scales and follow the same distribution as stable cold dark matter at super-horizon scales, decaying cold dark matter have the opposite behaviour: At small scales it follows its stable counterpart, while at large and even super-horizon scales it exhibit more structure. At the perturbation level we might then consider any difference between stable and decaying cold dark matter to be a general relativistic correction. Combining all three matter species into a single component, corresponding to ignoring the additional terms in \eqref{eq:continuity_combined} and \eqref{eq:euler_combined} --- transforming them back to \eqref{eq:continuity_dcdm_newv} and \eqref{eq:euler_dcdm_newv} or equivalently \eqref{eq:drift_op_final} and \eqref{eq:kick_op_final} --- is then acceptable when running simulations in small boxes or when otherwise ignoring GR effects.

Ignoring the $\bigl(\delta_{\text{M}}^{\text{Nb}} - \delta_{\text{dcdm}}^{\text{Nb}}\bigr)$ and $\bigl(\vec{v}_{\text{M}}^{\text{Nb}} - \vec{v}_{\text{dcdm}}^{\text{Nb}}\bigr)$ terms in \eqref{eq:continuity_combined} and \eqref{eq:euler_combined}, the evolution of the combined matter component is then governed by the same drift and kick operators as each individual matter species, \eqref{eq:drift_op_final} and \eqref{eq:kick_op_final}, with $\text{m}=\text{M}$. The mass reduction captured by the effective equation of state $\mathscr{w}_{\text{m}}$ is shown in Fig.~\ref{fig:w_eff}. All plotted lines appear very similar, with $\mathscr{w}_{\text{m}}$ moving closer to the stable matter limit of $\mathscr{w}_{\text{m}}=0$ for lower values of $\Gamma_{\text{dcdm}}$. The only deviant behaviour seen is that of $\mathscr{w}_{\text{M}}$ for $\Gamma_{\text{dcdm}}=\SI{100}{km.s^{-1}.Mpc^{-1}}$ at late times, where we see a sudden drop. For all models we have $\mathscr{w}_{\text{M}}(a\rightarrow \infty)=0$ as eventually only stable matter remains. For the large value of $\Gamma_{\text{dcdm}}=\SI{100}{km.s^{-1}.Mpc^{-1}}$ most of the initial decaying cold dark matter has already decayed away at $a=1$, placing the maximum point of $\mathscr{w}_{\text{M}}$ in the past.

The complete implementation of decaying cold dark matter --- including a modified \CLASS{} version, $N$-body mass reduction, dark radiation, the GR correction potentials $\gamma^{\text{Nb}}$ and $\gamma^{\text{Nb}}_{\text{lapse}}$ as well as the option to combine matter species --- is available\footnote{\url{https://github.com/jmd-dk/concept}} as of \CONCEPT{} version 0.3.0.

\section{Numerical setup and results}

\subsection{Simulation parameters}\label{subsec:sim_params}

In order to test the implementation outlined in the previous section we perform a range of $N$-body simulations using the publicly available \CONCEPT{} $N$-body solver \cite{Dakin:2017idt}. All \CONCEPT{} simulations in this work use cosmological parameters as listed in table~\ref{table:class_parameters} and a neutrino sector of three massless neutrinos. The \CONCEPT{} simulations all begin at $a=0.01$ and use potential grids (all of $\phi_{\text{sim}}$, $\phi_{\text{GR}}$ and $\gamma_{\text{lapse}}^{\text{Nb}}$) of size $1024^3$ and similarly have $N=1024^3$ particles. All \CONCEPT{} simulations are carried out in box sizes of $(65536\,\text{Mpc}/h)^3$, $(8192\,\text{Mpc}/h)^3$ and $(1024\,\text{Mpc}/h)^3$, the power spectra from which are patched together to give the ones presented.

We seek to vary both the amount of decaying cold dark matter $\Omega_{\text{dcdm}}$ as well as the decay rate $\Gamma_{\text{dcdm}}$ across the simulation suite. However, the current fractional energy density of decaying cold dark matter $\Omega_{\text{dcdm}}$ depends explicitly on $\Gamma_{\text{dcdm}}$, and so these are not independent parameters. To this end, we define $\widetilde{\Omega}_{\text{dcdm}}$ to be the current fractional energy density that the decaying cold dark matter component \emph{would} have, had $\Gamma_{\text{dcdm}}=0$. In this case, the energy density would scale like the usual $a^{-3}$ and so $\widetilde{\Omega}_{\text{dcdm}}$ may be written as\footnote{An alternative interpretation of the specification \eqref{Omega_dcdm_tilde} is that of a primordial fractional energy density, in the sense that the ratio of decaying to stable cold dark matter really is $\widetilde{\Omega}_{\text{dcdm}}/\Omega_{\text{cdm}}$ at early times, where the background densities of both species have identical scalings $a^{-3}$. This is how $\widetilde{\Omega}_{\text{dcdm}}$ is viewed in \cite{Audren:2014bca}, where it is called $\Omega^{\text{ini}}_{\text{dcdm}}$.}
\begin{equation}
	\widetilde{\Omega}_{\text{dcdm}} = \rho_{\text{crit},0}^{-1}\!\lim_{\phantom{^+}a\rightarrow 0^+} a^3\bar{\rho}_{\text{dcdm}}(a) \,. \label{Omega_dcdm_tilde}
\end{equation}
The parameters varied between the simulations are then the (primordial) fraction of the total cold dark matter energy content that is constituted by decaying cold dark matter,
\begin{equation}
    f_{\text{dcdm}} \equiv \frac{\widetilde{\Omega}_{\text{dcdm}}}{\Omega_{\text{cdm}}+\widetilde{\Omega}_{\text{dcdm}}}\,,
\end{equation}
as well as the decay rate $\Gamma_{\text{dcdm}}$. The values used for the simulation suite can be found on the upper and right perimeter of Fig.~\ref{fig:relative}.

\begin{table}[tb]
    \begin{center} 
        \begin{tabular}{l c} 
            \hline
            Parameter & Value  \\
            \hline
            $A_\text{s}$  & $\phantom{\mathclap{\raisebox{10pt}{.}}}2.1 \times 10^{-9}$ \\
            $n_\text{s}$ & $0.96$ \\
            $\tau_\text{reio}$ & $0.0925$  \\
            $\Omega_{\text{b}}$ & $0.049$  \\
            $\Omega_{\text{cdm}} + \widetilde{\Omega}_{\text{dcdm}}$ & $0.27$ \\
            $h$ & $0.67$ \\
            \hline						
        \end{tabular}
    \end{center}
    \caption{Cosmological parameters common to all \CLASS{} and \CONCEPT{} runs used.}
    \label{table:class_parameters} 
\end{table}

From CMB data only \cite{Poulin:2016nat} found a fairly robust upper bound on the product $f_{\text{dcdm}} \Gamma_{\text{dcdm}}$ of $f_{\text{dcdm}} \Gamma_{\text{dcdm}} < \SI{6.3e-3}{Gyr^{-1}} = \SI{6.2}{km.s^{-1}.Mpc^{-1}}$. However, they also found that for certain combinations of data the bound can loosen to  $f_{\text{dcdm}} \Gamma_{\text{dcdm}} < \SI{15.5}{km.s^{-1}.Mpc^{-1}}$.
Using the CMB bound we see that all models in the first column and the upper two models of the second column of Fig.~\ref{fig:relative} are allowed by data, while the rest is disallowed. Using the looser bound, only the lower two models of the third column are disallowed.

%\begin{table}[tb]
%    \begin{center} 
%        \begin{tabular}{l c c} 
%            \hline
%            Simulation & $f_{\text{dcdm}}$ & $\phantom{\mathclap{\Big(}}\Gamma_{\text{dcdm}}\,\bigl[\si{km.s^{-1}.Mpc^{-1}}\bigr]$  \\
%            \hline
%            A & 0.0    &    $\phantom{00}0$  \\
%            B & 0.1    &    $\phantom{00}1$  \\
%            C & 0.1    &    $\phantom{0}10$  \\
%            D & 0.1    &    $\phantom{}100$  \\
%            E & 0.3    &    $\phantom{00}1$  \\
%            F & 0.3    &    $\phantom{0}10$  \\
%            G & 0.3    &    $\phantom{}100$  \\
%            H & 1.0    &    $\phantom{00}1$  \\
%            I & 1.0    &    $\phantom{0}10$  \\
%            J & 1.0    &    $\phantom{}100$  \\
%            \hline						
%        \end{tabular}
%    \end{center}
%    \caption{Cosmological parameters of decaying dark matter used for the simulations. {\bf Not sure if we need this table}}
%    \label{table:sims} 
%\end{table}

\subsection{Results}

\begin{figure}[t]
\begin{center}
\hspace*{-0.03\textwidth}\includegraphics[width=1.06\textwidth]{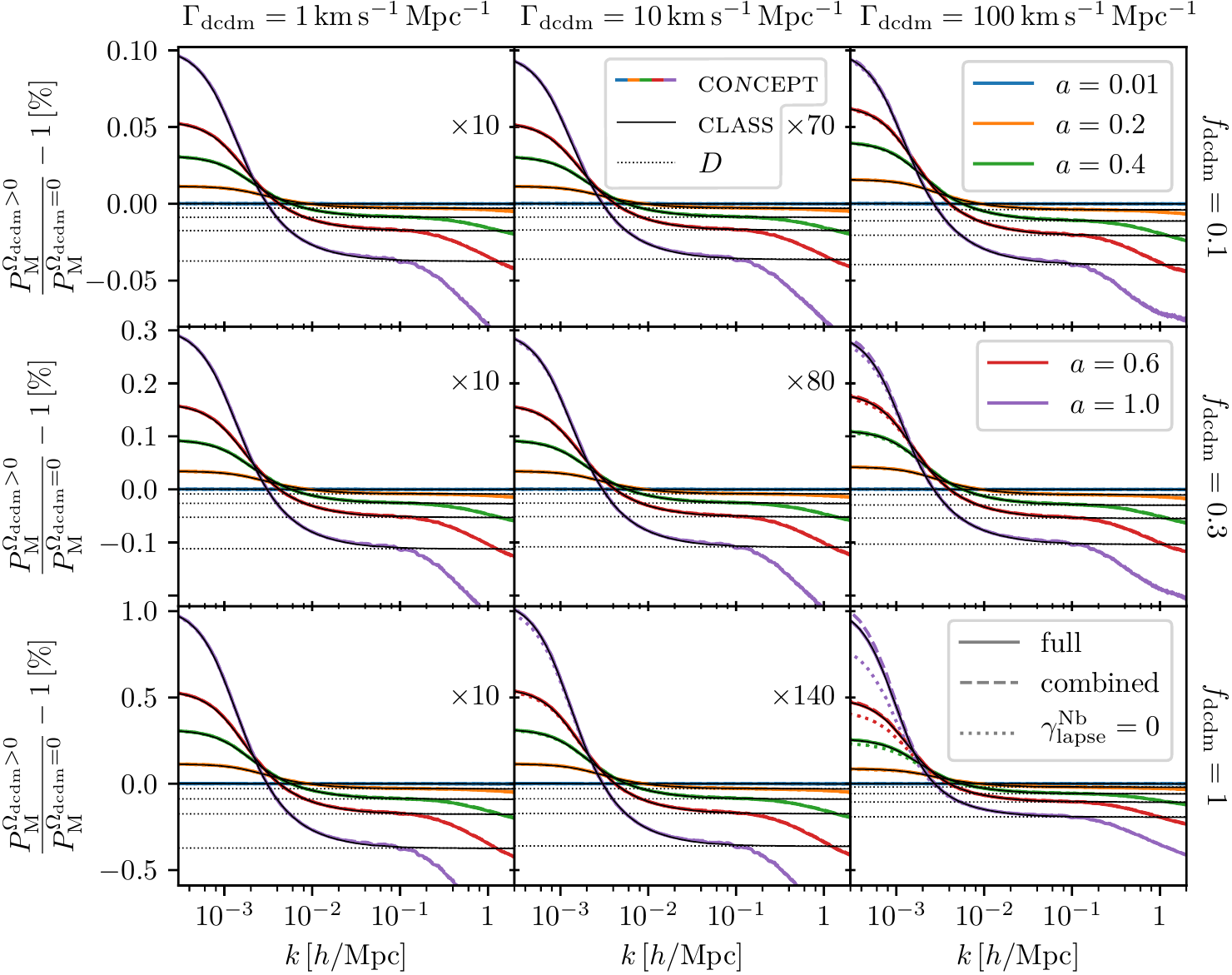}
\end{center}
\caption{Relative total matter power spectra between models with and without decaying dark matter. Coloured lines show results from \CONCEPT{} simulations, whereas full black lines indicate the corresponding linear results from \CLASS{}, all in $N$-body gauge. Each coloured line is present in three versions; solid and dotted lines result from simulations where decaying and stable matter are solved as separate components, whereas dashed lines result from simulations with a single, total matter component. Furthermore the lapse force has been neglected for the dotted coloured lines. Finally, the horizontal dotted lines show the relative power as predicted by the linear growth factor $D$. The vertical axes are shared across each row, but for panels in the second and third column scaled according to the factor given to the left of their respective vertical axes.
\label{fig:relative}}
\end{figure}

Results from our suite of $N$-body simulation are shown in Fig.~\ref{fig:relative}, in which we plot the matter power spectra of dcdm models relative to the benchmark $\Lambda$CDM model. Qualitatively, all models exhibit the same behaviour: There is an overall lowering of power originating in the changed background expansion rate and the diminishing mass in the matter component. This effect can be calculated by solving the Newtonian perturbation equations with the modified background and are shown as horizontal dotted lines in Fig.~\ref{fig:relative} and corresponds to the effect one would see when incorporating dcdm in a Newtonian $N$-body code.
For $k \lesssim 10^{-2} \, h/{\rm Mpc}$ power in the dcdm model rises rapidly and becomes larger than in the $\Lambda$CDM model. This effect comes from the addition of the dark radiation component, as well as the accompanying modification to the metric potential $\gamma^{\text{Nb}}$ (see e.g.\ \cite{Audren:2014bca} for a discussion of this)\footnote{As discussed in \cite{Audren:2014bca}, the rapid potential variation on these large scales also leads to a substantially enhanced late-ISW effect, which is perhaps the primary observational signature of the decaying cold dark matter scenario.}.

\begin{figure}[t]
\begin{center}
\includegraphics[width=0.85\textwidth]{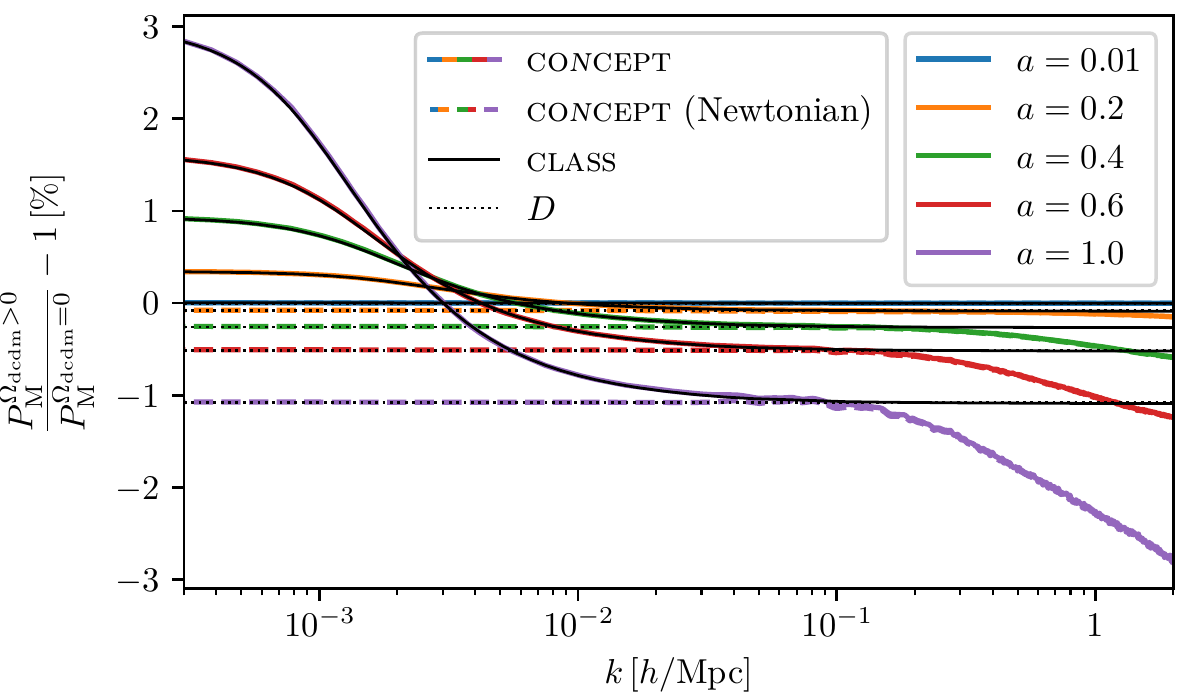}
\end{center}
\caption{Relative total matter power spectra between the model $\{f_{\text{dcdm}}=0.3,\,\Gamma_{\text{dcdm}}=\SI{10}{km.s^{-1}.Mpc^{-1}} \}$ and $\Lambda$CDM. Solid coloured lines, solid black lines and dotted black lines show results from full \CONCEPT{} simulations, \CLASS{} computations and the Newtonian growth factor $D$, respectively, and are identical to the ones presented in the center panel of Fig.~\ref{fig:relative}. The dashed coloured lines result from Newtonian \CONCEPT{} simulations, i.e.\ simulations with all GR effects neglected so that the only force felt by the particles is that of Newtonian self-gravity.
\label{fig:newtonian}}
\end{figure}

Though only clearly visible for the models with large $f_{\text{dcdm}}$ and $\Gamma_{\text{dcdm}}$, all \CONCEPT{} power spectra in Fig.~\ref{fig:relative} are presented in three versions, produced from slightly different simulations; 1) full simulations where $\text{b}+\text{cdm}$ and dcdm are simulated as two separate components each consisting of $N$ particles, 2) simulations where all matter is simulated as a combined $\text{b}+\text{cdm}+\text{dcdm}$ component (still using the now slightly erroneous \eqref{eq:drift_op_final} and \eqref{eq:kick_op_final}) and 3) simulations with separate $\text{b}+\text{cdm}$ and dcdm components but where the lapse potential $\gamma_{\text{lapse}}^{\text{Nb}}$ is neglected. While the full simulations always exactly match the \CLASS{} solutions at large scales, the others fail to match at very large scales for large values of $f_{\text{dcdm}}$ and $\Gamma_{\text{dcdm}}$. The errors obtained by collecting together all three matter species in a single component, corresponding to approximating $\bigl(\delta_{\text{M}}^{\text{Nb}} - \delta_{\text{dcdm}}^{\text{Nb}}\bigr) = \bigl|\vec{v}_{\text{M}}^{\text{Nb}} - \vec{v}_{\text{dcdm}}^{\text{Nb}}\bigr|=0$ in \eqref{eq:continuity_combined} and \eqref{eq:euler_combined}, are small even for $f_{\text{dcdm}}=1$ and our most extreme value of $\Gamma_{\text{dcdm}}=\SI{100}{km.s^{-1}.Mpc^{-1}}$. It is then not just tempting but generally perfectly allowable to save on computational resources and simulate all matter using a single component. A similar verdict may be passed on the approximation $\gamma_{\text{lapse}}^{\text{Nb}}=0$, though the errors produced here is significantly larger for large $f_{\text{dcdm}}$. In the case of $\{f_{\text{dcdm}}=1,\,\Gamma_{\text{dcdm}}=\SI{10}{km.s^{-1}.Mpc^{-1}}\}$, combining all matter species leads to no visible error, while neglecting $\gamma_{\text{lapse}}^{\text{Nb}}$ produces small but noticeable errors.

For intermediate and large values of $k$ the two approximate schemes agree with the full simulations. For large values of $k$ we see the usual non-linear suppression dip (see e.g.\ \cite{Brandbyge:2008rv}) arising when comparing a model with a smaller amount of linear power to the benchmark $\Lambda$CDM model.

In Fig.~\ref{fig:newtonian} we again show the relative power between a model with and without decaying cold dark matter, but this time we additionally show results from \CONCEPT{} simulations that are purely Newtonian, i.e.\ ones that have $\phi_{\text{GR}}=\gamma_{\text{lapse}}^{\text{Nb}}=0$. We obtain the exact same non-linear small-scale solution as with the full GR simulations, but now the large-scale behaviour matches that of the linear Newtonian growth factor $D$, not general relativistic linear perturbation theory (\CLASS{}). We compute $D$ through the usual second-order ODE
\begin{equation}
    \ddot{D} + \mathcal{H}\dot{D} - 4\pi G a^2\bar{\rho}_{\text{M}}D = 0\,,
\end{equation}
with decaying dark matter included in $\bar{\rho}_{\text{M}}(\tau)$ as well as decaying dark matter and dark radiation included in $a(\tau)$ and $\mathcal{H}(\tau)$.

In the models studied here the separation between small $k$ where GR and radiation effects must be taken into account, and large $k$ where structures are non-linear is almost absent, contrary to e.g.\ models with massive neutrinos \cite{Tram:2018znz} or time-varying dark energy equation of state \cite{Dakin:2019vnj}. However, since Fig.~\ref{fig:newtonian} show that the matter power spectrum for large $k$ is unchanged when our simulations are restricted to be Newtonian, even in the present case we have the required scale separation needed for our approach of treating all species --- except for matter --- in pure linear theory.

%%%%%%%%%%%%%%%%%%%%%%%%%%%%%%%%%%%%%%%%%%%%%%%%%%%%%%%%%%%%%%%%%%%%
\section{Discussion}

Given our lack of knowledge of the nature of dark matter, it is of interest to study whether significant amounts of dark matter can have decayed before the present epoch.
This has been the subject of numerous studies in the literature, using a variety of different cosmological observables to test for signatures of dark matter decay.
In general, cosmological observations exclude the possibility that more than a small fraction of the cold dark matter can have decayed. However, an intriguing possibility might be that a small amount of dark matter decay might help alleviate some of the tensions between some locally measured cosmological parameters (i.e.\ $H_0$ and $\sigma_8$) and their values inferred from CMB measurements.

In this paper we studied the non-linear structure formation properties of models with decaying cold dark matter using a set-up which is fully consistent with GR at the linear perturbation theory level and at the same time follows the fully non-linear evolution on small scales.

For models which are not excluded by current data we find that on large scales, Newtonian simulations of decaying dark matter can be off by tens of percent relative to the full results including GR effects and radiation perturbations.
Our results agree at the sub-per-mille level with results from the linear theory Einstein-Boltzmann code \CLASS{} on these scales, and require the inclusion of several different effects not previously included in simulations of decaying dark matter.

However, we also find that these corrections are negligible on scales where structure starts to go non-linear at the present epoch, thus justifying our treatment of these effects using linear perturbation theory.
The separation of scales between the small $k$ range where GR and radiation corrections must be accounted for, and the intermediate to large $k$ range where non-linear effects are important also means that it should be possible to make working semi-analytic models of e.g.\ the non-linear matter power spectrum using methods such as HALOFIT \cite{Smith:2002dz}. 
Finally we note that although we specifically studied dark matter to dark radiation decay in this paper, the formalism we employ should also work for other models of dark matter to dark radiation conversion (such as those discussed in \cite{Bringmann:2018jpr}).

\section*{Acknowledgements}
This work was supported by the Villum Foundation.

\appendix

%%%%%%%%%%%%
\bibliographystyle{utcaps}

%\bibliography{refs}
\providecommand{\href}[2]{#2}\begingroup\raggedright
\endgroup

\end{document}